\documentclass[journal]{IEEEtran}
\usepackage{cite}
\usepackage{amsmath,amssymb,amsfonts}
\usepackage{algorithm}
\usepackage{algorithmic}
\usepackage{graphicx}
\usepackage{epstopdf}
\usepackage{textcomp}
\usepackage{array}
\usepackage{multirow}
\usepackage{color}
\usepackage{url}
\usepackage{subfigure}
\usepackage{setspace}
\usepackage{hyperref}
\usepackage[hyphenbreaks]{breakurl}
\begin{document}
\title{Open DNN Box by Power Side-Channel Attack}
\author{
	\vskip 1em
	Yun~Xiang,
	Zhuangzhi~Chen,
    Zuohui Chen,
    Zebin~Fang,
    Haiyang~Hao,
   Jinyin~Chen,
	Yi~Liu, \emph{Member,~IEEE,}
   Zhefu~Wu,
   Qi~Xuan, \emph{Member,~IEEE}
   and~Xiaoniu Yang
	}

\maketitle
\begin{abstract}
Deep neural networks are becoming popular and important assets of many AI companies. However, recent studies indicate that they are also vulnerable to adversarial attacks. Adversarial attacks can be either white-box or black-box. The white-box attacks assume full knowledge of the models while the black-box ones assume none. In general, revealing more internal information can enable much more powerful and efficient attacks. However, in most real-world applications, the internal information of embedded AI devices is unavailable, i.e., they are black-box. Therefore, in this work, we propose a side-channel information based technique to reveal the internal information of black-box models. Specifically, we have made the following contributions: (1) we are the first to use side-channel information to reveal internal network architecture in embedded devices; (2) we are the first to construct models for internal parameter estimation; and (3) we validate our methods on real-world devices and applications. The experimental results show that our method can achieve 96.50\% accuracy on average. Such results suggest that we should pay strong attention to the security problem of many AI applications, and further propose corresponding defensive strategies in the future.
\end{abstract}

\begin{IEEEkeywords}
Deep learning, machine learning, model identification, side-channel attack, adversarial attacks.
\end{IEEEkeywords}


\section{Introduction}
\label{sec:introduction}
\IEEEPARstart{R}{ecently}, deep neutral networks (DNNs) have been the focus of research and widely used in many artificial intelligence (AI) related areas~\cite{lecun2015deep}, such as image classification, object detection, video recognition, and natural language processing etc. Many DNNs are deployed and implemented on embedded devices, e.g., robots, self-driving cars, and smartphones etc, or to solve specific industrial tasks like pearl classification~\cite{xuan2018automatic,8575147}, fault diagnosis~\cite{8114247,Liu_2019,chen2019multiview} and soft sensors~\cite{liu2017flame,liu2018ensemble}. With the miniaturization of DNNs and the development of AI chips~\cite{howard2017mobilenets,sandler2018inverted}, DNNs on embedded hardware are becoming increasingly common and meanwhile, vulnerable to various attacks. In this work, we demonstrate, for the first time, a powerful side-channel attack (SCA) based technique to the embedded DNN devices which has the potential to reveal critical internal information of DNNs.


Currently, it was found that DNNs are vulnerable to adversarial attacks in the form of tiny perturbations~\cite{szegedy2013intriguing}. In other words, by adding small controllable noises to the input, we can mislead the network to generate incorrect results. This poses a security problem for the application of DNN, e.g., attacks on road signs~\cite{DBLP:journals/corr/EvtimovEFKLPRS17} may bring a safety hazard to autonomous driving; attacks on face attributes~\cite{rozsa2017facial,mirjalili2017soft} may invalidate many face recognition applications; and attacks on robot vision~\cite{melis2017deep} may challenge the application of robots. The mainstream attacking models include black-box and white-box attacks~\cite{akhtar2018threat}. The difference lies at the structural knowledge of the network. The black-box attack assumes no prior knowledge of the neural network model~\cite{su2017one,tramer2017space}. While the white-box attack, on the other hand, relies upon the complete structure information, including both network architecture and parameter values~\cite{goodfellow6572explaining,kurakin2016adversarial,DBLP:journals/corr/abs-1812-01713}. Apparently, the white-box is much more powerful and resilient than the black-box model at the cost of feasibility, i.e., most of the embedded devices are considered as black-box. To leverage this tradeoff and significantly improve the attack performance, it is required to obtain additional neural network information for the embedded hardware.

SCA is a powerful tool to obtain hardware information~\cite{kocher1996timing, kocher1999differential}. It takes advantage of side-channel signals, such as power consumption, computing time, and electromagnetic radiation etc., to reveal hidden information inside the embedded hardware~\cite{zhou2005side}. Traditionally, it is used to extract the secretive keys during the encryption/decryption process. However, since during the computation process of the DNNs, the side channel information shows strong correlations to the network structure and its parameters, we envision that SCA can be used for embedded AI devices and reveal their network architectures and even the corresponding parameters. In other words, we intend to use SCA to open the black-box of DNNs, which can facilitate adversarial attacks by transforming a black-box attack to an at least partial white-box, or gray-box, attack. Note that, as a Chinese aphorism says ``\emph{the law of defending a city is born from siege; if you want to defend well, you must attack well}", we hope this study of SCA on DNNs could provide useful insights for researchers to propose defensive strategies in the future to better protect AI devices.

In this work, we propose a SCA based technique to explore the network structure and parameter properties, which might be used for subsequent attacks. First, we collect the device power consumption traces. Then we use machine learning techniques to identify the structure and parameter values. Our technique is based on the assumption that the AI device employs existent architectures and pre-trained parameters, which is appropriate for many real-world applications~\cite{singla2016food,schwarz2015rgb,keraspretrained}. Specifically, we have made the following contributions.

\begin{enumerate}
\item We are the first to employ the SCA method to identify the DNN structures in embedded devices.
\item We are also the first to use SCA to estimate the sparsity of DNN parameters. Based on the sparsity features, we derive the pre-trained parameter values.
\item We validate the effectiveness of our techniques on a real-world embedded platform.
\end{enumerate}
To perform SCA, we use the power reading measurement during DNN processing. The experiment results show that our technique can achieve more than 96.5\% attacking success rate.

The remaining parts of this paper is organized as follows. Section~\ref{sec:related work} introduces the related works. Section~\ref{sec:preliminary} describes basic knowledge of DNN structures and parameters. Section~\ref{sec:method} presents the underlying theories and details of our technique. Section~\ref{sec:experiments} shows the experiment setup and results. Section~\ref{sec:conclusion} concludes the paper.

\section{Related Work}
\label{sec:related work}
The current work is related to deep neural networks, embedded AI hardware, and side channel attacks, which will be briefly reviewed.

\subsection{Deep neural networks}
In this work, we aim to identify the internal DNN architecture~\cite{bengio2009learning, schmidhuber2015deep}. In the real-world applications, several DNN architectures are widely used. For example, AlexNet~\cite{krizhevsky2012imagenet} is popular for its success in the 2012 ImageNet competition~\cite{russakovsky2015imagenet}. GoogleNet~\cite{szegedy2015going} significantly increases the depth of DNN. ResNet~\cite{he2016deep} beats human experts in image recognition. VGGNet~\cite{simonyan2014very} and RCNN~\cite{girshick2014rich} are widely used for their breakthrough in object detection. There are also networks specific to mobile applications~\cite{howard2017mobilenets,sandler2018inverted}. Currently, most engineers design their AI product based on the exist architectures. Therefore, by identifying the existent popular architectures, we expect to be able to break a large portion of such AI products.

\subsection{Embedded AI hardware}
Embedded AI devices are becoming more and more popular recently, and many AI specific embedded products are announced. Cambricon developed a series of embedded AI hardware based on the Diannao architecture~\cite{chen2014dadiannao, zhang2016cambricon}. They can be used as AI processing units and handle jobs such as face recognition and object detection etc. Intel proposed the Movidius Myriad X Visual Processing Unit (VPU), a deep learning based visual AI core to accelerate applications such as drones, smart cameras, and VR/AR helmets, etc. Huawei released Ascend series of AI processing units which are designed for full range scenarios. Nvidia and Facebook announced plans for DNN processing units. Google also designed dedicated hardware Tensor Processing Units (TPU) for its AI services.

In this work, we develop a raspberry pi based experimental platform. As an Arm cortex based system, Raspberry Pi shares the common architecture of many existent devices. Therefore, the experiments performed on this system can be easily migrated to other similar systems.

\subsection{Side-channel attack}
Side-channel attack (SCA) is a very powerful tool in attacking encrypted systems. Traditionally, the encryption process is considered as a perfect black-box. However, in real-world applications, information can be leaking~\cite{zhou2005side}.  Initially, SCA is focused on differential power analysis~\cite{kocher1999differential} and timing attacks~\cite{kocher1996timing}. Later, more side-channel information and attacking methods are developed. Yuval et al.~\cite{yarom2014flush+} propose a cache based SCA to extract the private encryption keys. Genkin et al.~\cite{genkin2014rsa} extract full 4096-bit RSA keys successfully using the computer audio information during the decryption process. By cloning the USIM cards, Liu et al.~\cite{liu2015small} can recover the encryption key and other information contained from the 3G/4G USIM cards. Defense against SCA is also well studied~\cite{tiri2005vlsi}.

Naturally, this powerful attacking method can be applied to reveal DNN architectures or some related information. Duddu et al.~\cite{DBLP:journals/corr/abs-1812-11720} used timing side channels to infer the depth of the network; Batina et al.~\cite{DBLP:journals/corr/abs-1810-09076} show that side channel attacks can roughly obtain information on activation functions, number of network layers, number of neurons, number of output categories, and weights in the neural network; Another close work is to obtain the input image by analyzing the power trace in the first convolution layer~\cite{wei2018know}. So far, the current work is the first attempt trying to reveal the internal DNN architectures of embedded devices using power SCA.

\section{Preliminary}
\label{sec:preliminary}
In this section, we introduce the basics of DNN and the corresponding parameters.

\subsection{DNN architectures}
\begin{figure*} [!t]
\centering
\includegraphics[width=0.99\textwidth]{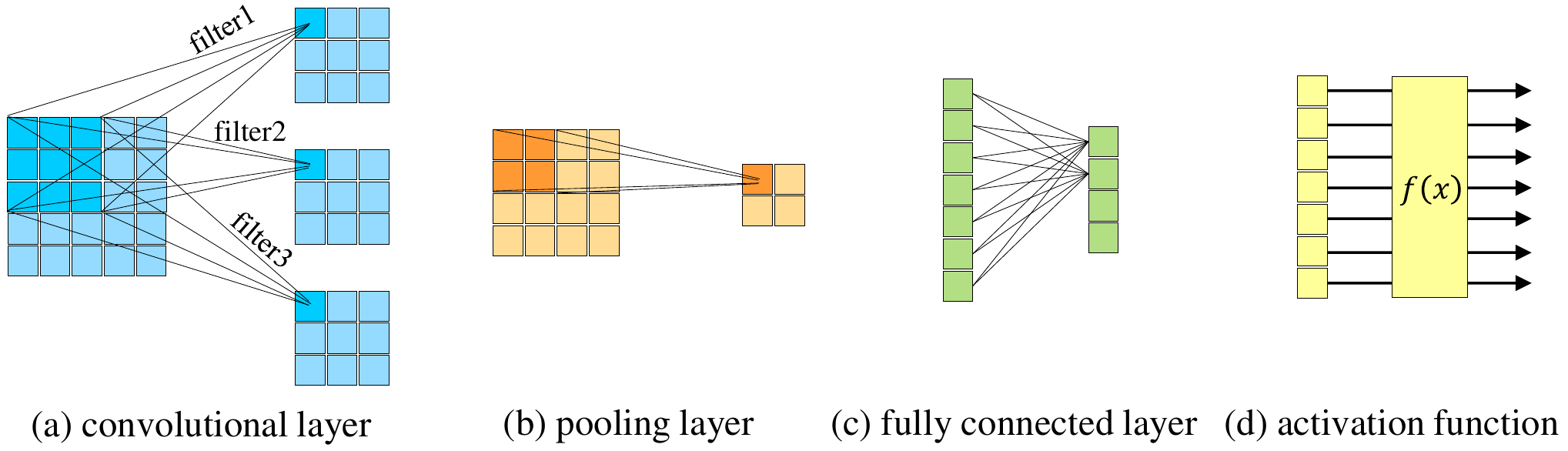}
\caption{Some typical components of DNNs. (a) The convolution layer performs a series of convolution operations on the image or feature maps by convolution kernels; (b) the pooling layer generally includes average pooling and maximum pooling; (c) the fully connected layer implements a fully connected linear operation of neurons; and (d) an activation function, which is always nonlinear, is added to the output of other typical neural network layers to make the model more powerful.}
\label{Fig:DNN}
\end{figure*}

The mainstream DNN architectures typically share some common and critical components, e.g., in visual applications, convolutional layers are usually used to extract features, while fully connected layers are used for classifications. In this paper, we mainly focus our research on the computer vision related DNN models. Fig.~\ref{Fig:DNN} shows the typical components, including convolutional layers, pooling layers, fully connected layers, and activation function. The convolution layers use various convolution kernels to filter the images. The pooling layer is essential to reduce the layer dimensions. The fully connected layers consist of  fully connected neurons. The activation function is typically a non-linear function added to the output of each neuron.

One critical observation is that different components require different computational cost. Therefore, different architectures have different power consumption patterns, which makes DNN architectures vulnerable to SCAs.

\subsection{Parameters and sparsity}
\label{sec3:B}

Parameters, or neuron weights and biases, define a DNN model. Typically, a DNN model is first trained using back propagation. During the training phase, the parameters are  continuously updated. Then in the inferring phase, the parameters combined are used to perform various classification operations. Training a DNN from scratch requires numerous computation resources and time~\cite{srivastava2015training,ioffe2015batch,krizhevsky2012imagenet}. Therefore, in real-world applications, people usually derive the DNN based on pre-trained parameters on existent models.

Another problem for embedded AI applications is that the computational resources on embedded platforms are very limited. To address this problem, various parameter pruning techniques were proposed. Han et al.~\cite{han2015deep} deeply compressed neural networks by pruning and trained quantization and Huffman coding and~\cite{han2016eie} proposed an energy efficient inference engine, which infers the compressed network model and acceleration vector multiplication by weight sharing. Yu et al.~\cite{yu2017nisp} proposed the Neuron Importance Score Propagation algorithm to better reduce redundant connections. Kang et al.~\cite{kang2018accelerator} proposed a pruning scheme for convolutional neural networks (CNN) running on accelerators. Lin et al.~\cite{lin2018accelerating} proposed a novel global and dynamic pruning scheme to prune redundant filters for CNN acceleration. The basic idea of parameter pruning is to set some unimportant weights to be zero. Therefore, the computational cost is reduced while the performance of the whole network retains. Parameter sparsity is defined as the proportion of zero-valued parameters. Obviously, for different DNN models with the same architecture, power consumptions can vary significantly because of different parameter sparsity. In that case, other than the DNN architecture, it is possible that the device power traces can be used to reveal the actual weights of a large portion of pre-trained neurons.

\section{Side channel power based technique}
\label{sec:method}
In this section, we present our technique in detail. Specifically, we first develop power consumption theories and models on various DNN layers. Then, we discuss the impact of parameter sparsity on DNN power cosumptions. Finally, we describe our SCA based technique.

\subsection{DNN power models}
First, we build power computation models for each kind of neural network layer.

\subsubsection{Convolutional layer}
As shown in Fig~.\ref{Fig:DNN} (a), convolutional layer typically consists of many filters and is used to extract features at different level of abstractions. Specifically, the convolutional operation can be described by
\begin{equation}
\label{eq:conv}
a_{i,j} = \sum^{C-1}_{d=0}\sum^{F-1}_{m=0}\sum^{F-1}_{n=0}\omega _{m,n}x_{i+m,j+n}+\omega_b,
\end{equation}
where $a_{i,j}$ represents the output of a single convolutional filter, $x_{i,j}$ represents the input value, $C$ represents the number of input channels, $F$ represents the filter kernel size, and $\omega_{m,n}$ and $\omega_b$  represent the parameter weights and bias, respectively. Therefore, according to Eq.~(\ref{eq:conv}), the total operation number of a single convolutional filter is calculated by
\begin{equation}
\begin{cases}
N_{mul} = C\times F\times F\\
N_{add} = C\times F\times F,\\
\end{cases}
\end{equation}
where $N_{mul}$ is the number of multiplication operations and $N_{add}$ is the number of addition operations. Assuming that the input size is $C\times L\times W$, the filter stride is $S$, and the number of convolutional kernels is $N$, the total operation number of the convolutional layers is calculated by
\begin{equation}
\begin{split}
N_{cov} & = f_{cov}(C,L,W,S,N,F)\\
& = \begin{cases}
N_{mul} = (\frac{L}{S}\times\frac{W}{S})\times(N\times C\times F\times F)\\
N_{add} =(\frac{L}{S}\times\frac{W}{S})\times (N\times C\times F\times F).\\
\end{cases}
\end{split}
\end{equation}

Thus, the power consumption of a single convolutional layer can be derived as
\begin{equation}
P_{conv}(C,L,W,S,N,F) = \frac{p_mLWNCF^2}{S^2} + \frac{p_aLWNCF^2}{S^2}, \label{power_conv}
\end{equation}
where $p_m$ and $p_a$ are the average multiplication and addition operation power consumptions, respectively, of the device. It is observed that the power consumption of the convolution layer is strongly correlated to the input and kernel sizes, i.e., the convolution layer architectures.

\subsubsection{Pooling layer}
To reduce the growing feature dimensions without hurting the performance, there is typically a pooling layer between two consecutive convolutional layers, as shown in Fig.~\ref{Fig:DNN} (b). Without loss of generalization, we employ the maximum pooling method here. Thus, the total number of operations for the pooling layer is
\begin{equation}
N_{pl} = f_{pl}(C,L,W,S,F) = C\times \frac{L}{S}\times \frac{W}{S}\times F\times F.
\end{equation}
where $(C,L,W)$ is the input size, $S$ is the pooling stride, and $F$ is the pooling window size, for simplicity. Therefore, the power consumption of a pooling layer is calculated by
\begin{equation}
P_{pl}(C,L,W,S,F) = \frac{p_cCLWF^2}{S^2} \label{power_pl},
\end{equation}
where $p_c$ represents the average comparison operation power consumption of the device. Eq.~(\ref{power_pl}) indicates that the pooling layer power consumption is also largely determined by the DNN architecture.

\subsubsection{Fully connected layer}
After the convolution layer and pooling layer, DNNs typically use fully connected layers, as shown in Fig.~\ref{Fig:DNN} (c), to process the extracted features. The fully connected layers consists of several layers of fully connected neurons. The operation number of a single fully connected layer can be calculated by
\begin{equation}
N_{fc} = f_{fc}(X,Y) = \begin{cases}
N_{mul} = X\times Y\\
N_{add} =X\times Y,\\
\end{cases}
\end{equation}
where $X$ is the number of input neurons and $Y$ is the number of output neurons. Thus, the power of fully connected layer can be derived as
\begin{equation}
P_{fc}(X,Y) = p_mXY + p_aXY.\label{power_fc}
\end{equation}

\subsubsection{Activate function}
For each neuron, its value should be judged by an activation function, as shown in Fig.~\ref{Fig:DNN} (d). There are many different types of activation functions, such as relu, tanh, sigmod, and softmax, etc.  The total number of operations for the activation function is calculated by
\begin{equation}
compute_{ac} = f_{ac}(C,L,W) = \alpha\times C\times L\times W,
\end{equation}
where $\alpha$ is the operational coefficient which is determined by the specific type of the activation function. Therefore, the power consumption of the activate function can be derived as
\begin{equation}
P_{ac}(C,L,W) = p_{ac}\alpha CLW \label{power_ac},
\end{equation}
where $p_{ac}$ is the power consumed by one operation in activation function. In general, the power consumption of the activation function is linear to the inputs and relatively small compared to other layers.

\subsubsection{Overall power consumption}
\begin{figure} [!t]
\centering
\includegraphics[width=0.49\textwidth]{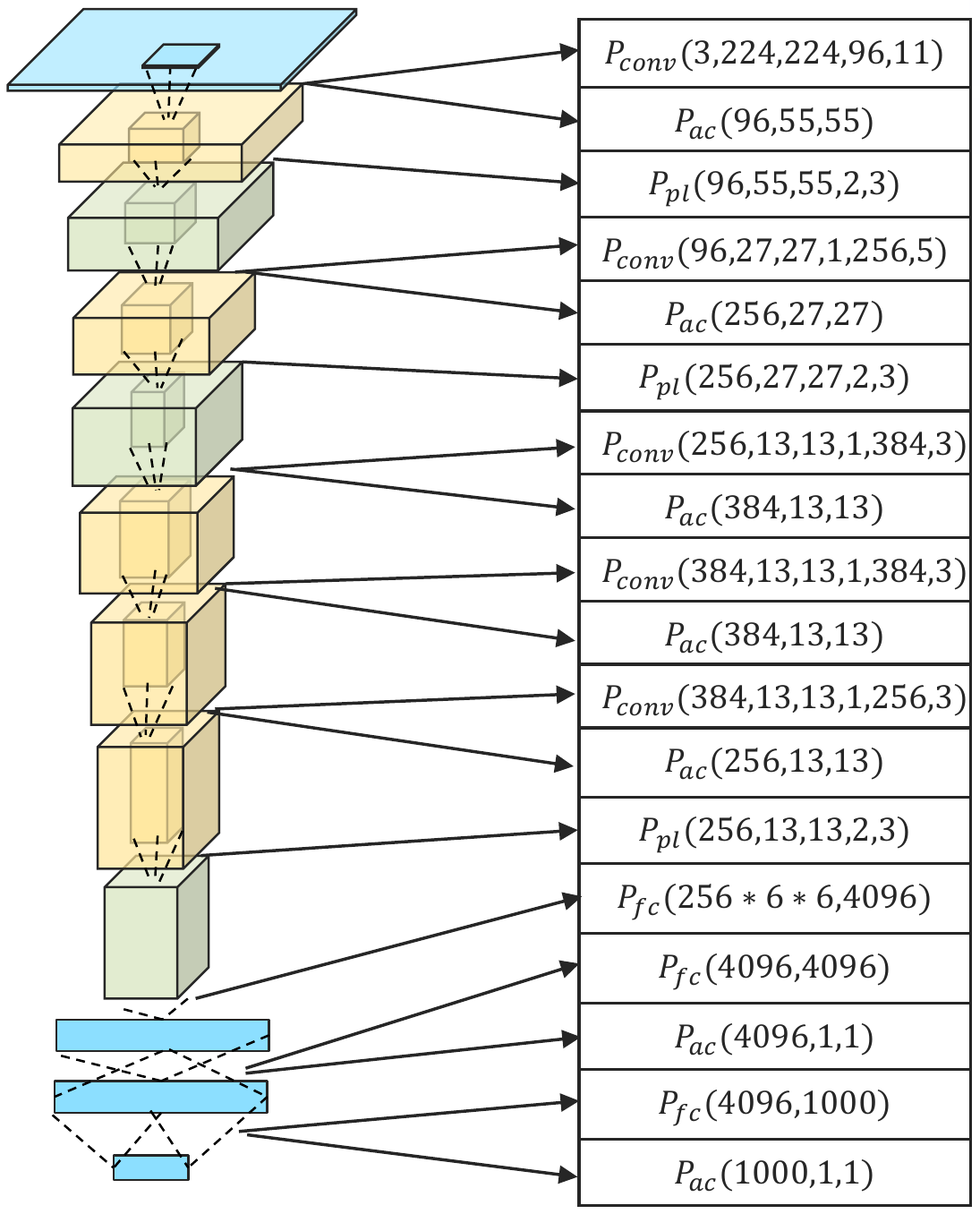}
\caption{The power model of the Alexnet. The Alexnet consists of five convolutional layers, three pooling layers, three fully connected layers, and some common activation functions. 
}
\label{Fig:alexnet-power}
\end{figure}

So far, we have built the power consumption models for the major components of modern DNNs. There are many other special operations. However, their computational cost is usually negligible. Therefore, in this work, we construct our general power model based on Eq.бл(\ref{power_conv}), (\ref{power_pl}), (\ref{power_fc}), and (\ref{power_ac}).  Fig.~\ref{Fig:alexnet-power} shows a simple example of our model. This simple DNN consists of five convolutional layers, three pooling layers, three fully connected layers, and common activation functions. The operation of each layer can be represented by our corresponding power model.

It is observed that the variation of DNN architecture can have significant impact on the power consumptions. Thus, theoretically we can identify the specific architecture by analyzing and classifying the operational power traces.

\subsection{Parameter sparsity model}
\begin{figure*} [!t]
\centering
\includegraphics[width=0.99\textwidth]{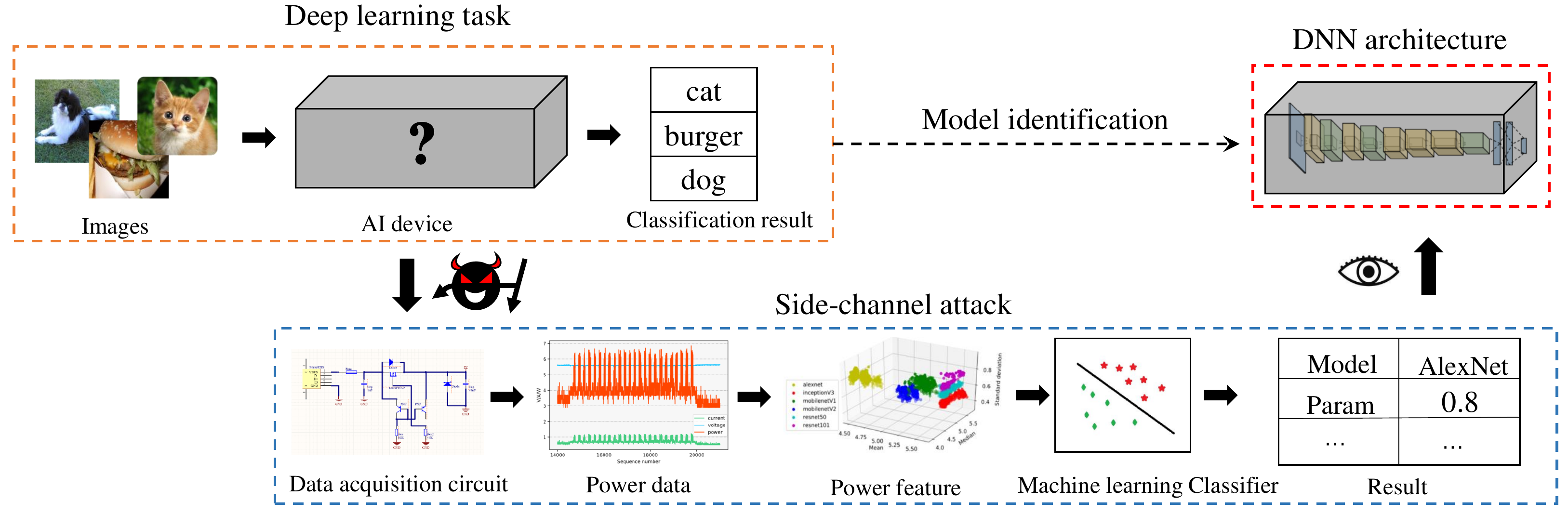}
\caption{The framework of our SCA on DNN models. We collect voltage and current data while the AI device is running a model, get the power features, and then establish the classifier to identify the DNN structure and sparsity, so as to realize the attack.
}
\label{Fig:overall}
\end{figure*}

In the previous section, we have generalized the models to identify the DNN architectures. In order to further improve the efficiency of adversarial attacks on DNN, it is desirable to derive the actual parameters, such as weights and biases, of each neuron. However, unlike the architecture, the power variation of different parameter values is very insignificant, unless it is zero.  it is observed that many AI models use at least partial pre-trained parameters. Moreover, some advanced AI chips have hardware-level parameter pruning to improve the computational efficiency while maintaining the performance. Combining the pre-training and parameter pruning techniques, it is possible to further identify the actual parameter values.

It is assumed that the pre-training and pruning techniques for various models are known to the attackers. Thus, during the operation of the AI device, different pre-trained parameter sets with various pruning method can generate unique sparsity in the neurons, which can lead to differentiable power traces. Parameter pruning mainly occurs in the convolution and the fully connected layers. The power consumption of the convolution layer and that of the fully connected layer are thus changed to
\begin{equation}
P^{'}_{conv}(C,L,W,S,N,F) = \lambda _1 P_{conv}(C,L,W,S,N,F),
\end{equation}
\begin{equation}
P^{'}_{fc}(X,Y) = \lambda _2 P_{fc}(X,Y),
\end{equation}
where $\lambda _1, \lambda _2 \in (0,1]$ are the parameter sparsity coefficients of the convolution layer and the fully connected layer, respectively. When no pruning operation is applied, the values of $\lambda$ is simply set to 1. Different pre-training parameters combined with different parameter pruning techniques can lead to significant power variations. Thus, this observation can be used to identify the actual parameter values of DNN model in the AI device, which makes it even more vulnerable.

\subsection{General power model}
Based on the above DNN architecture and parameter power models, we build our general power model to classify the specific architecture and the corresponding parameter values. It includes training and testing phases.

The framework of our SCA on DNN models is shown in Fig~\ref{Fig:overall}. We continuously collect voltage and current data while the AI device is running a model, calculate the power and power features to form a power-feature data set. We then randomly choose some data to train a classifier, and input the rest of the power-feature data into the trained classifier to get the model structure and the parameter sparsity. More specifically, for each model, we sample $n$ pairs of current and voltage data (different DNNs have different $n$, for the different time they take to calculate a single image) and then calculate the average, median, and standard deviation of the power as the power features. For each model, we repeat the above sampling and data processing many times to form a power-feature data set. Finally, we design classifier $D$ and use the data set for training. To further evaluate the parameters, we set different sparsity for each model, so the final labels include the model structure $y$ and parameter sparsity $\lambda$.

Our SCA is summarized as Algorithm~\ref{alg:1}, where we use the classifier $D$, voltage array $U$, and  current array $I$ as the inputs, and then use $D$ to classify and generate the outputs.

\begin{algorithm}[!h]
    \setstretch{1.25}
    \renewcommand{\algorithmicrequire}{\textbf{Input:}}
	\renewcommand{\algorithmicensure}{\textbf{Output:}}
    \caption{Side-channel attack (SCA)}
    \label{alg:1}
    \begin{algorithmic}[1]
        \REQUIRE
        Voltage array $U=\{u_1,u_2,...,u_n\}$ and current array $I=\{i_1,i_2,...,i_n\}$ obtained by data acquisition card; The trained classifier $D$.
        \ENSURE
        Scalars $y$ and $\lambda$ representing the prediction results of the DNN structure and sparsity, respectively.
        \STATE Calculate power: $P=U*I=\{p_1,p_2,...,p_n\}$;

        \STATE Obtain the power features:\\
        $p_{mea}=\sum_{i=1}^n{p_i}/n$;\\
        $p_{mid}=sort(P)[n/2]$;\\
        $p_{std}=\sqrt{\sum_{i=1}^n(p_i-p_{mea})^2/n}$.

        \STATE Enter ${p_{mea},p_{mid},p_{std}}$ into the classifier $D$ to get the output result:\\
        $y,\lambda=D(p_{mea},p_{mid},p_{std})$.

        \STATE Return $y,\lambda$.
    \end{algorithmic}
\end{algorithm}

\section{Experiments}
\label{sec:experiments}
In this section, we describe the experiments to validate the effectiveness of our techniques.

\subsection{Experiment setup}



\begin{figure} [!t]
\centering
\includegraphics[width=0.45\textwidth]{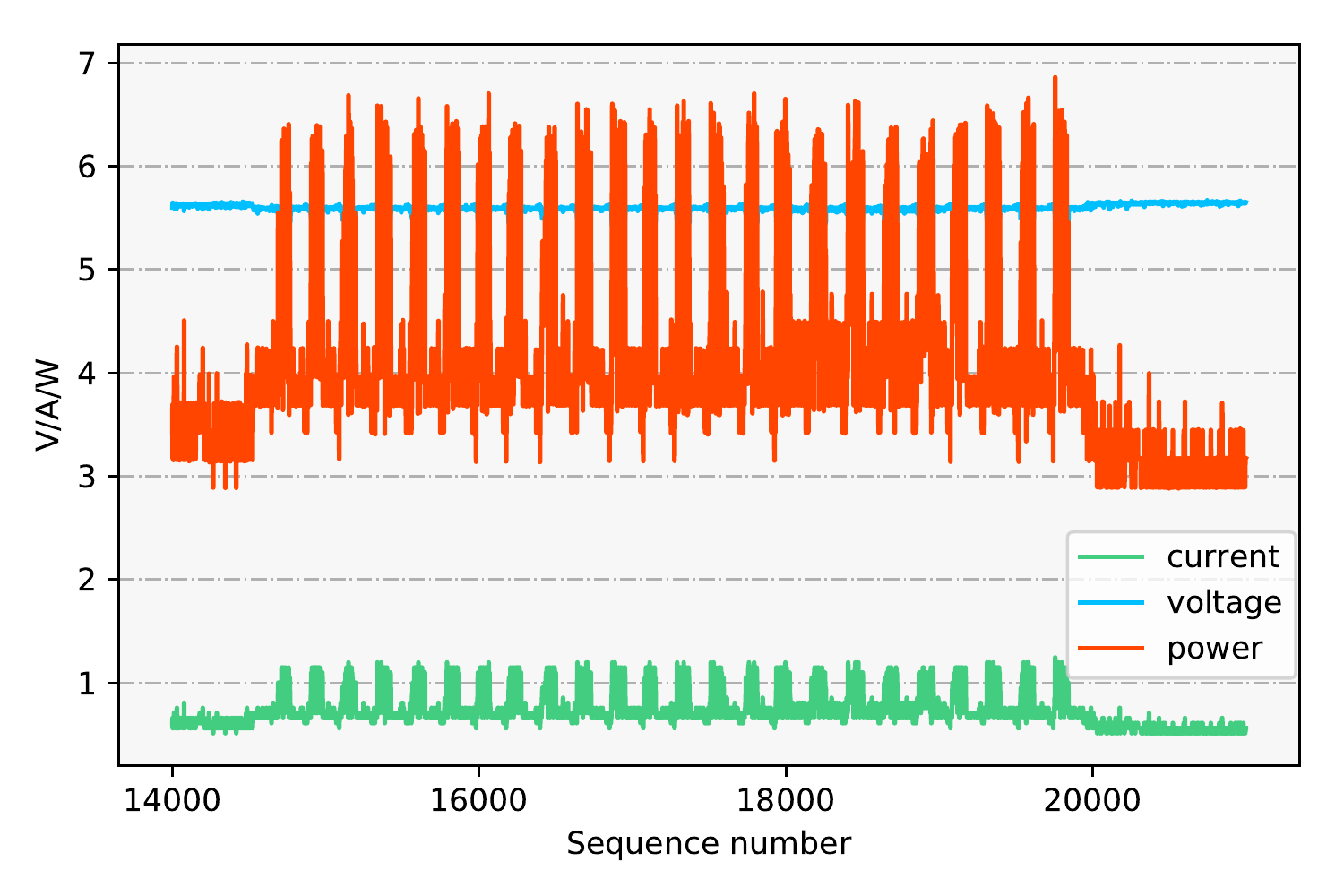}
\caption{The trace of using Alexnet to classify an epoch images (24 photos) on the Raspberry Pi. The classification of 24 images corresponds to the 24 peaks.}
\label{Fig:powertrace}
\end{figure}

To extract valid power data, we use an external data acquisition card running at 400$Hz$. Fig.~\ref{Fig:powertrace} shows an example of our data collection setup.

In this example, we use Alexnet for image classification and continuously input 24 images (an epoch).  The processing of the DNN requires a large amount of computing resources. Therefore, the device power increases significantly.  We can observe 24 peaks, corresponding to 24 images processed. Note that there is a start and end phase when the device runs the DNN model. Therefore, we remove those low-power phases and take the middle part of the data (about $14800 to 20000$ sequence number in the figure) as the input. Due to the instantaneous nature of the power, in the process of inference, we divide per five-images' inference into a group, take all the power data, calculate their mean, median and variance as the power-feature data of the model.

\begin{figure} [!t]
\centering
\includegraphics[width=0.45\textwidth]{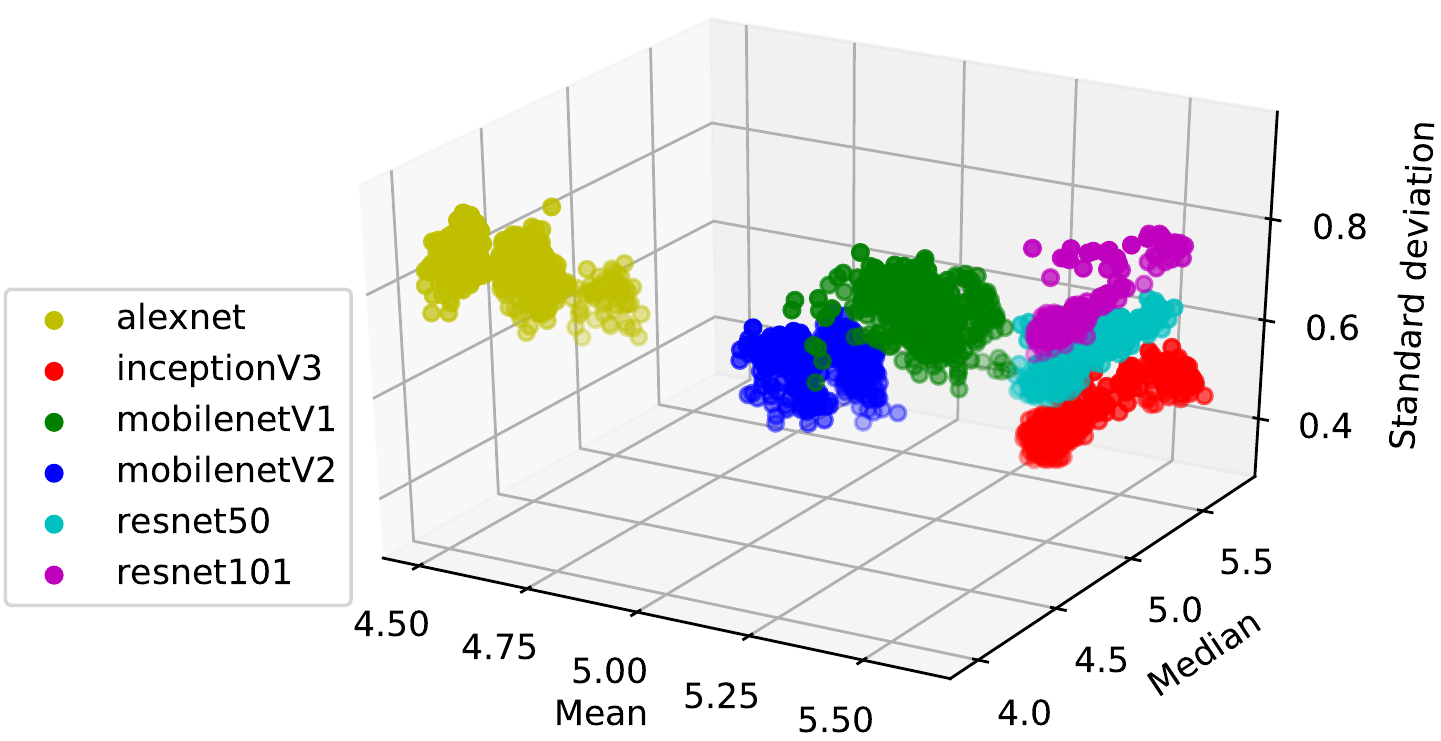}
\caption{Visualization of the power-feature data set. It can be seen that the power features of different DNN models have a high degree of discrimination.}
\label{Fig:vis}
\end{figure}

We implement six common DNN models on the Raspberry Pi and test them with the same set of images. After data acquisition and processing, we obtain a power-feature data set, divided into 6 categories, as shown in Fig.~\ref{Fig:vis}.



\subsection{Architecture identification}
\begin{figure} [!t]
\centering
\includegraphics[width=0.45\textwidth]{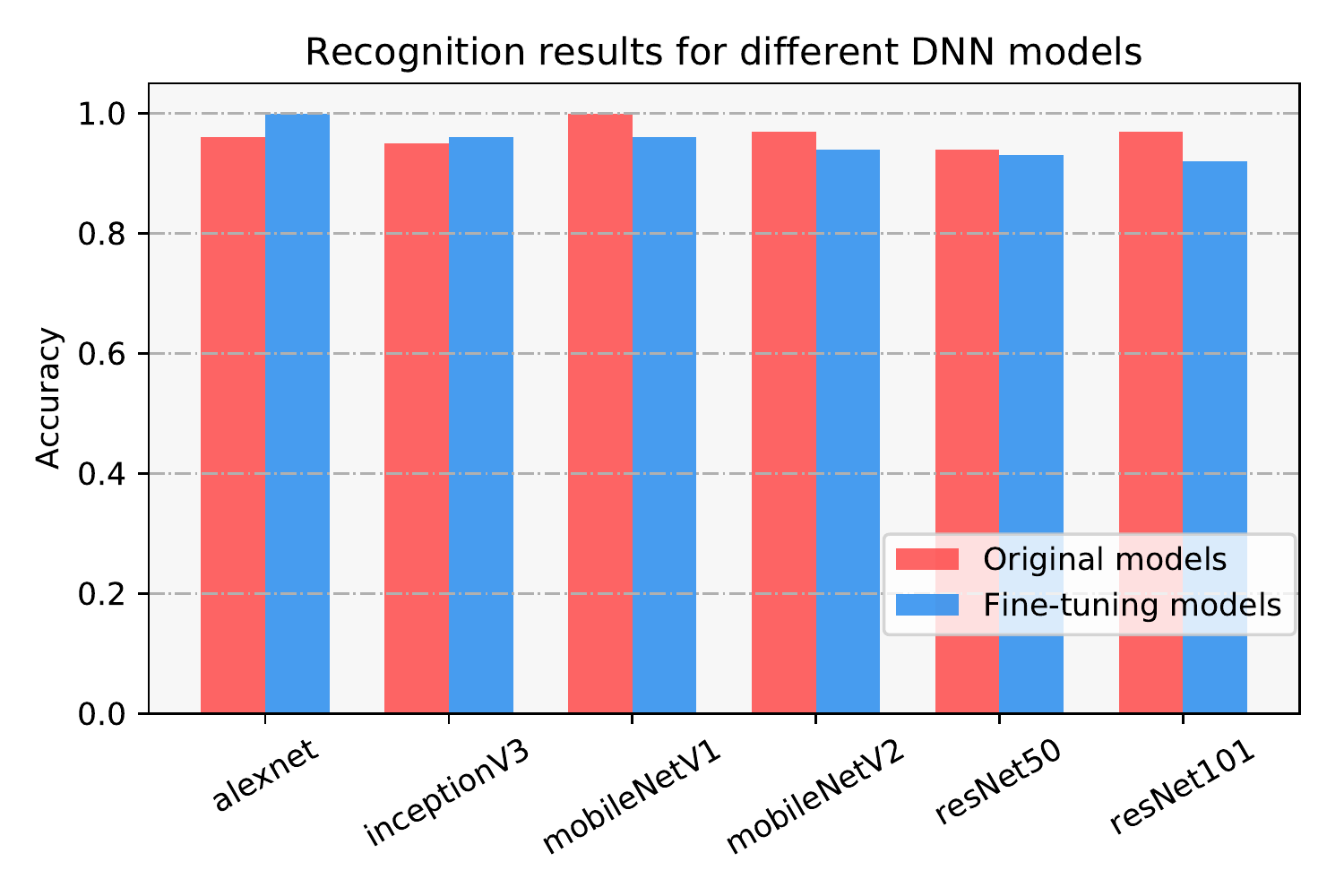}
\caption{Recognition results for different DNN models. From red bars, it can be seen that with the power features, the original models can be well identified by using the SVM algorithm, achieving an average accuracy of 96.50\%. From blue bars, it can been seen that after adding the fine-tuning part, the power characteristics of different models still have a high degree of discrimination, achieving an average accuracy of 95.17\% .}
\label{Fig:result1}
\end{figure}

In this work, we use machine learning techniques to identify the DNN architectures. First, we randomly divide the power-feature data set into training set and testing set with a ratio of 4:1. Here, we simply employ the widely used SVM classifier. The results on test set are shown in Fig.~\ref{Fig:result1}, where the red bars is the architecture identification accuracies. The average classification accuracy reaches 96.50\%. It should be noted that we do not intend to compare the performance of different machine learning algorithms, since this work focus on proposing the overall SCA framework and SVM already achieves quite high accuracy. By using more advanced algorithms, the experimental results could be further improved.

\begin{table}[!t]
	\caption{The fine-tuning strategy for the DNN models.}
	\label{finetune}
    \centering
	\begin{tabular}{c|c|c}
		\hline\hline
        Models & Fine-tuning strategy & Hyperparam\tabularnewline
        \hline
        \multirow{4}{*}{Alexnet}
        &\multirow{4}{*}{\shortstack{Change the number of \\the last three layers}} & 4096*4096*1000 \tabularnewline
		& & 4096*4096*500 \tabularnewline
        & & 4096*2048*500 \tabularnewline
        & & 2048*2048*500 \tabularnewline
        \hline
        \multirow{4}{*}{InceptionV3}
        &\multirow{4}{*}{\shortstack{Change the number of \\the output layer}} & 1000 \tabularnewline
		& & 750 \tabularnewline
        & & 500 \tabularnewline
        & & 250 \tabularnewline
        \hline
        \multirow{4}{*}{Resnet50}
        &\multirow{4}{*}{\shortstack{Change the number of \\the output layer \\and the different \\building blocks~\cite{he2016deep}}} & 1000,(3,4,6,3) \tabularnewline
		& & 1000,(3,6,4,3) \tabularnewline
        & & 500,(3,6,4,3) \tabularnewline
        & & 500,(3,5,5,3) \tabularnewline
        \hline
        \multirow{4}{*}{Resnet101}
        &\multirow{4}{*}{\shortstack{Change the number of \\the output layer \\and the different \\building blocks}} & 1000,(3,4,23,3) \tabularnewline
		& & 1000,(3,7,20,3) \tabularnewline
        & & 500,(3,14,13,3) \tabularnewline
        & & 500,(3,17,10,3) \tabularnewline
        \hline
        \multirow{4}{*}{MobilenetV1}
        &\multirow{4}{*}{\shortstack{Change the number of \\the output layer}} & 1000 \tabularnewline
		& & 750 \tabularnewline
        & & 500 \tabularnewline
        & & 250 \tabularnewline
        \hline
        \multirow{4}{*}{MobilenetV2}
        &\multirow{4}{*}{\shortstack{Change the number of \\the output layer}} & 1000 \tabularnewline
		& & 750 \tabularnewline
        & & 500 \tabularnewline
        & & 250 \tabularnewline
        \hline
        \hline
	\end{tabular}
\end{table}

In the previous experiment setup, we assume that the pre-trained DNNs are deployed exactly unchanged. However, this may not be true in the real-world applications. In general, many DNN models can be divided into two parts. The first part is for the feature extraction, such as the convolutional layers and pooling layers. The second part is for the classification, such as the fully connected layers. In real-world applications, the feature extraction part is typically with little or no change while on the other hand, the classification part is always fine-tuned based on the actual application.

Therefore, to further explore the performance of our technique with the changing classification layers, we repeat the experiment with changing fully connected layer hype-parameters.  We employ three fine-tuning methods for each original architecture, as shown in Table~\ref{finetune}, with each architecture containing four entries. The first one is the original structure and the others are the fine-tuned ones. The power traces from all the fined-tuned architectures are mixed with the original ones. Then the architecture identification process is repeated on the new data set. The classification results for fine-tuning models are still shown in Fig.~\ref{Fig:result1} (blue bars). The average performance is slightly reduced to 95.17\%, which demonstrates that our SCA is quite robust even with the fine-tuning of the classification layers.

\begin{figure*} [!t]
\centering
\includegraphics[width=0.9\textwidth]{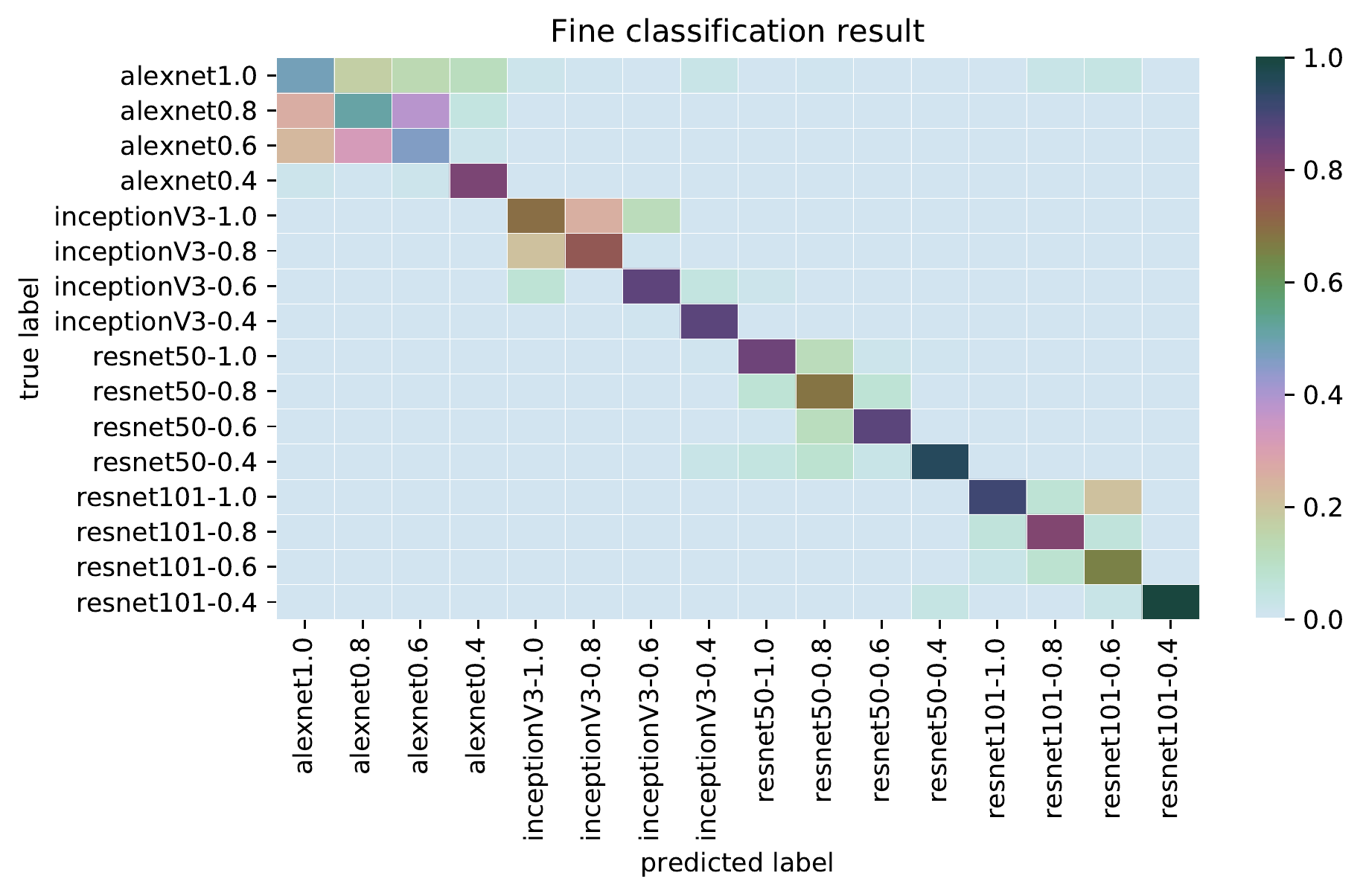}
\caption{The confusion matrix for the fine classification of DNN models. As we can see, the recognition accuracy of parameter sparsity is lower than that of the model structure. The average accuracy of the fine classification is 75.88\%.}
\label{Fig:fine-cla}
\end{figure*}

\subsection{Parameter evaluation}
\label{Parameter evaluation}
The evaluation of the model parameters is also important. For example, the traditional white-box attack requires full knowledge of the DNN model, including parameter values. However, estimating the model parameters are usually quite challenging. In this work, we assume that  pre-trained architectures and partial model parameters could be employed in certain real-world applications. Moreover, due to the techniques such as parameter tunings, different models could have varying sparsity, i.e., the ratio of zeros. We thus use the model sparsity as features to infer the model parameters.

Specifically, we employ dropout~\cite{DBLP:journals/corr/abs-1207-0580}  in all the convolutional and fully connected layers and set different dropout scales to derive various parameter sparsity. The dropout scale is set to 1.0, 0.8, 0.6, and 0.4, respectively. In this experiment, we select four DNN models. The classification results of the same model with varying parameter sparsity are shown in Table~\ref{sparsity}. The average identification rate is about 76.38\%. The Alex network is relatively hard to identify since it is relatively small and consumes less power. In general, in the large networks, the accuracy is well above 82\%.

\begin{table}[!t]
	\caption{Accuracy of the same model in different parameter sparsity (dropout scales).}
	\label{sparsity}
    \centering
	\begin{tabular}{c|c|c|c}
		\hline\hline
        Models & Parameter sparsity & Accuracy & Average\tabularnewline
        \hline
        \multirow{4}{*}{Alexnet}
        & 1.0 & 88\% & \multirow{4}{*}{54.25\%}\tabularnewline
		& 0.8 & 36\% & \tabularnewline
        & 0.6 & 44\% & \tabularnewline
        & 0.4 & 49\% & \tabularnewline
        \hline
        \multirow{4}{*}{InceptionV3}
        & 1.0 & 94\% & \multirow{4}{*}{82.75\%}\tabularnewline
		& 0.8 & 86\% & \tabularnewline
        & 0.6 & 76\% & \tabularnewline
        & 0.4 & 75\% & \tabularnewline
        \hline
        \multirow{4}{*}{Resnet50}
        & 1.0 & 100\% & \multirow{4}{*}{84.00\%}\tabularnewline
		& 0.8 & 84\% & \tabularnewline
        & 0.6 & 65\% & \tabularnewline
        & 0.4 & 87\% & \tabularnewline
        \hline
        \multirow{4}{*}{Resnet101}
        & 1.0 & 100\% & \multirow{4}{*}{84.5\%}\tabularnewline
		& 0.8 & 64\% & \tabularnewline
        & 0.6 & 88\% & \tabularnewline
        & 0.4 & 86\% & \tabularnewline
        \hline
        \hline
	\end{tabular}
\end{table}

We have performed model architecture and model parameter estimation. However, these two processes may affect each other. For example, the varying model sparsity may lead to wrong identification of the model architecture. Therefore, we need to check the general model identification results which include both varying architecture and varying parameters.

Thus, by mixing up the architecture and parameter sparsity variations, we have  a total of 16 different categories in the data set. The experimental results are shown in Fig~\ref{Fig:fine-cla}. The confusion matrix demonstrates that, even with varying parameter sparsity, we can still get the structure of a DNN model with relatively high accuracy, i.e., more than 95\%. The accuracy of parameter sparsity recognition under the fine classification task is 75.88\%, which is reduced a little but generally acceptable.

\section{Conclusion}
\label{sec:conclusion}
In this paper, we propose a side-channel attack (SCA) method to reveal the internal structure and model parameters for DNN models.  We design a raspberry pi based platform to derive the power signature of embedded AI devices, and then use machine learning algorithms to identify the specific DNN architectures. Moreover, we differ the parameter sparsity to model the pre-training of DNNs.  In general, our technique can identify both the architecture and model parameters with quite high accuracy, indicating that we should pay strong attention to the security problem of many AI applications.

In the future, we  will first improve the experimental platform by considering more diverse DNN architectures and parameters; then, we will use more advanced machine learning algorithms to identify DNN models more precisely; finally, we will try to propose defensive strategies to protect the model information from SCA.


\bibliographystyle{Bibliography/IEEEtranTIE}
\bibliography{BIB_17-TIE-2179}\ 

\end{document}